\documentclass[aps,prl,superscriptaddress,twocolumn, secnumarabic, amssymb, floatfix, 10pt]{revtex4-2}

\usepackage[T1]{fontenc}
\usepackage[nolist]{acronym}
\usepackage{microtype}
\usepackage{graphicx} 
\usepackage{lipsum} 
\usepackage[english]{babel}
\usepackage{amsmath}
\usepackage{hyperref}
\usepackage[capitalize]{cleveref}
\usepackage{cancel}
\usepackage{bm}
\usepackage{multirow}
\usepackage{colortbl,hhline}
\usepackage{siunitx}
\usepackage{color}
\usepackage[percent]{overpic}

\usepackage{blindtext}
\usepackage{xr}

\newcommand{\ttu}{\tau^*}
\newcommand{\tg}{\tau}

\newcommand{\td}{\tau_{D}}

\newcommand{\tm}{\tau_{m}}

\newcommand{\tmfp}{\tau_{\mathrm{MFP}}}

\newcommand{\UPPMax}{U''_{\mathrm{max}}}

\newcommand{\OmegaMax}{\omega_{\mathrm{max}}}

\newcommand{\ptpx}{p(\mathrm{TP}|x)}
\newcommand{\ptpxMax}{\mathrm{max}[p(\mathrm{TP}|x)]}

\DeclareMathOperator\erf{erf}

\graphicspath{{./figs/}}

\begin{document}

\title{Non-Markovianity increases transition path probability}

\author{Florian N. Br\"unig}
\affiliation{Department of Physics, Freie Universität Berlin, 14195 Berlin, Germany}

\author{Benjamin A. Dalton}
\affiliation{Department of Physics, Freie Universität Berlin, 14195 Berlin, Germany}

\author{Jan O. Daldrop}
\affiliation{Department of Physics, Freie Universität Berlin, 14195 Berlin, Germany}

\author{Roland R. Netz}
\email[]{rnetz@physik.fu-berlin.de}
\affiliation{Department of Physics, Freie Universität Berlin, 14195 Berlin, Germany}

\date{\today}

\begin{abstract}
  Defining low-dimensional reaction coordinates is crucial for analyzing the dynamics of complex systems and for comparison with experiments. The maximal value of the transition-path probability along the reaction coordinate $x$, $\ptpx$, is a common estimator for reaction-coordinate quality by comparing to the theoretical maximal value of 1/2 in the overdamped Markovian limit. We show by analytical arguments and simulations that for non-Markovian dynamics $\ptpx$ is non-monotonic as a function of the memory time and exceeds 1/2 for long memory time. This disqualifies $\ptpx$ as a criterion for reaction coordinate quality.
\end{abstract}


\maketitle

\begin{acronym}[Bash]
   \acro{GH}{Grote and Hynes}
   \acro{GLE}{generalized Langevin equation}
   \acro{LE}{Langevin equation}
   \acro{MD}{molecular dynamics}
   \acro{MSD}{mean squared displacement}
   \acro{PTP}{$p(\text{TP}|x)$}
   \acro{PMF}{potential of mean force}
   \acro{RC}{reaction coordinate}
   \acro{SM}{Supplementary Material}
   \acro{TP}{transition path}
   \acro{TS}{transition state}
\end{acronym}

Reaction coordinates are low-dimensional projections from the full phase space of many-body systems that are used to analyze the dynamics of chemical or biophysical reactions.
Whereas in experimental setups the reaction coordinate is usually defined by design of the experiment \cite{Zijlstra2020, Dudko2011, Truex2015, Neupane2012}, in the analysis of simulation data, where the full phase space is accessible, low-dimensional representations assist insightful interpretation
\cite{Frederickx2014}. 
Methods for finding and optimizing reaction coordinates are an active field of research \cite{Peters2016,Mehdi2022,Jung2019,Johnson2012}.
Because the description of a chemical reaction depends, in both experiment and computer simulation, on the reaction coordinate, it is important to define and develop methods to estimate
the `quality' of reaction coordinates. 

\Acp{TP} are trajectories in phase space that connect the reactant and product regions of a reaction directly without intermediate return into either region \cite{Orland2011}. 
In high-dimensional systems and at finite temperature, like for chemical reactions in solution, \acp{TP} form a diverse ensemble \cite{Kim2015, Daldrop2016, Carlon2018, Louwerse2022}. 
The widely-applied conditional transition-path probability, $\ptpx$, was introduced as a measure for reaction-coordinate quality by quantifying how well a reaction coordinate projects out the \ac{TP} ensemble \cite{Hummer2004}.
It is based on the commitor, or $p$-fold, analysis, that was developed in the context of \ac{TP} sampling~\cite{Dellago1998, Bolhuis2002, Allen2005, Sega2007}.
The $\ptpx$ method has been applied to \ac{MD} data for testing~\cite{Hansen2011, Hinczewski2010, Neupane2016, Zijlstra2020} and finding or optimizing reaction coordinates~\cite{Best2005, Dudko2011, Johnson2012, Peters2006, Peters2010a}.

The $\ptpx$ method was introduced for overdamped Markovian dynamics, in which case a value close to the theoretical maximum 1/2 indicates a good reaction coordinate \cite{Hummer2004,Peters2010b}.
Recently, the importance of inertial and non-Markovian memory effects have been demonstrated for chemical reactions and protein folding \cite{Lee2019,Ayaz2021,Brunig2022e,Dalton2023,Dalton2024, Dalton2024a}. 
Non-Markovian memory is produced by projection of the dynamics on low-dimensional reaction coordinates \cite{Ayaz2022}.
It clearly is important to analyze how the widely used $\ptpx$ method performs in the presence of inertial and non-Markovian effects.
More so, since it was already suggested that for inertial dynamics the maximal value of $\ptpx$ may increase beyond the benchmark value 1/2 for the overdamped scenario~\cite{Ballard2012,Berezhkovskii2018}. 
Non-Markovian effects on the other hand were previously argued to decrease the maximal value of $\ptpx$ \cite{Berezhkovskii2018}. 

We here analyze the behavior of $\ptpx$ for a one-dimensional reaction coordinate in the presence of inertial and non-Markovian memory friction effects. 
We find that the maximal value of $\ptpx$ may exceed 1/2 in the presence of either effect. Because of non-monotonicities, the benchmark value 1/2 is reached twice, for vanishing and long memory, and is therefore not a reliable indicator of reaction-coordinate quality or Markovianity. 
Using the transmission coefficient predicted by \ac{GH} theory \cite{Grote1980}, we give an analytical estimate for the maximal value of $\ptpx$, which serves as a reference for inertial or non-Markovian systems.

\begin{figure}[tbp]
  \centering
  \begin{overpic}[width=0.4\textwidth]{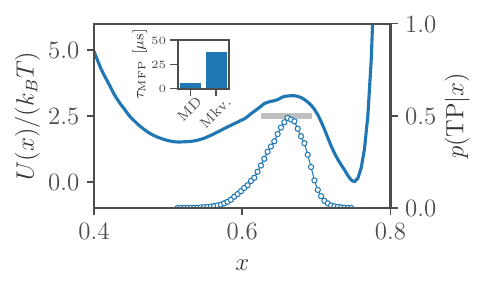}
  \end{overpic}
  \vspace{-4mm}
\caption{
Folding dynamics of the $\alpha$3D protein for the fraction of native contacts reaction coordinate $x$. The free-energy profile $U(x) = - k_B T \log[p(x)]$, where $p(x)$ is the probability density of $x$ and $k_BT$ the thermal energy, is shown as a solid blue line (left axis). The conditional probability to be on a transition path at position $x$, $\ptpx$, is shown with bullet markers (right axis), the short horizontal gray line indicates 1/2. In the inset, the mean first-passage time $\tmfp$ from the unfolded state to the barrier top obtained in \ac{MD} simulations is compared to an overdamped Markovian prediction, taken from \cite{Dalton2023}. 
Simulation data were obtained in \cite{Lindorff-Larsen2011}.
}
\label{ptpr_intro}
\end{figure}

\textbf{Analytical theory:}
The transition-path probability $\ptpx$ can be conveniently computed using the Bayesian equality $p(x|\text{TP})p(\text{TP})=p(x,\text{TP})=p(\text{TP}|x)p_{eq}(x)$  \cite{Hummer2004} (see \ac{SM} \ref{SI_method} \cite{suppMat}).
$p_{eq}(q)$ is the equilibrium distribution, $p(x|\text{TP})$ is the distribution of the transition-path probability along $x$ and thus $p(\text{TP}|x)$ the fraction of trajectories at $x$ that form \acp{TP}. In the overdamped Markovian limit, $\ptpx$ is directly related to the commitor probabilities $\phi_{A/B}(x)$ to reach the boundaries $A$ and $B$ \cite{Hummer2004} (see \ac{SM} \ref{SIptpx_od} and \ref{SIcommitor}).
\begin{align}
  \label{eq:ptp2}
  p(\text{TP}|x)=2\phi_A(x)\phi_B(x)=2\phi_A(x)(1-\phi_A(x)),
\end{align}
from which follows that $\ptpxMax=1/2$. 
This value was suggested as an estimator for reaction coordinate quality \cite{Hummer2004}.\\
In \cref{ptpr_intro}, $\ptpx$ of the prototypical fraction of native contacts reaction coordinate, here denoted as $x$, is shown for the folding dynamics of the protein $\alpha$3D as obtained from all-atom simulation data of the Shaw group and analyzed in our previous work \cite{Chung2015a,Dalton2023}.
The $\ptpx$ profile, shown with bullet markers, is narrow and clearly peaks with a value close to 1/2, spuriously indicating that $x$ is a good reaction coordinate \cite{Hummer2004} and that its dynamics is Markovian \cite{Berezhkovskii2018}. However, in the inset, a comparison of the mean first-passage time $\tmfp$ from \ac{MD} simulations with the overdamped Markovian prediction shows that the actual dynamics is significantly accelerated due to non-Markovian friction, in the line with the memory time $\tau = \SI{3.3}{\micro s}$ being  as large as the \ac{MD} folding time  \cite{Kappler2018, Ayaz2021, Dalton2023}. This example clearly indicates that using $\ptpx$ as indicator of Markovianity is problematic.\\
A versatile model for studying non-Markovian effects is the \ac{GLE} 
\begin{align}
m\ddot x (t) = &-\gamma_M \dot x(t) -\int_{0}^{t}\Gamma(t')\dot x(t-t')dt' \nonumber \\ 
&-  \nabla U[x(t)] + \eta(t),
\label{eq:gle}
\end{align}
which is derived by projection-operator techniques \cite{Mori1965a,Zwanzig1960,Ayaz2022}.
Here the friction memory kernel $\Gamma(t)$ acts in addition to an explicit Markovian friction with strength $\gamma_M$.  The random force $\eta(t)$ has zero mean $\langle \eta(t) \rangle = 0$ and 
$\langle \eta(t) \eta(t') \rangle =  k_BT [\Gamma(|t-t'|)+2\gamma_M \delta(t-t') ]$. The standard double-well potential, $U(x)=U_0(1-(x/L)^2)^2$, with barrier height $U_0$ is employed. Furthermore, as the simplest case, a single memory time scale $\tg$ is assumed, i.e. $\Gamma(t)=(\gamma_{nM}/\tg)\exp{(-t/\tg)}$, with $\gamma_{nM} = \int_0^{\infty} \Gamma(t) dt$.
The numerical implementation is detailed in \ac{SM} \ref{SIgle}. We introduce the diffusion time, $\tau_D=L^2\gamma/k_BT$, which is the average time for a freely diffusing particle to travel the distance $L$ in a flat potential landscape with friction constant $\gamma = \gamma_M + \gamma_{nM}$, and the inertial time scale $\tm=m/\gamma$ describing the transition from the inertial to the diffusive regime  \cite{Kappler2018,Brunig2022c}.

Note, that in the overdamped limit, $m\to0$, this model has previously been used to study the $\ptpx$ method \cite{Berezhkovskii2018}.
The present model not only allows to study inertial and non-Markovian effects in conjunction; $m\neq0$ is in fact required in the limit of purely non-Markovian friction, $\gamma_M \to 0$
as shown in \ac{SM} \ref{SIgle}.
It turns out that the results in the limit $\gamma_M \to 0$ differ significantly from the previous results \cite{Berezhkovskii2018}.

We first calculate $\ptpx$ in the inertial limit, $m \to \infty$, by assuming energy-conserving Newtonian dynamics; all configurations with a kinetic energy $E$ larger than the local potential energy relative to the barrier top $\Delta U(x)=U_0-U(x)$ exclusively form \acp{TP}. $\ptpx$ is thus given by integration over the kinetic energy distribution $p(E)=\exp(-E/k_BT)$,
\begin{align}
p&(\text{TP}|x)_{m \to \infty} = \int_{\Delta
U(x)}^{\infty}p(E)dE 
= 1- \erf\sqrt{\frac{\Delta U(x)}{k_BT}},
\label{eq:ptp2_inertial}
\end{align}
where $\erf$ denotes the error function. Thus $\ptpx=1$ at the barrier.

\begin{figure*}[tbp]
  \centering
\begin{overpic}[width=\textwidth]{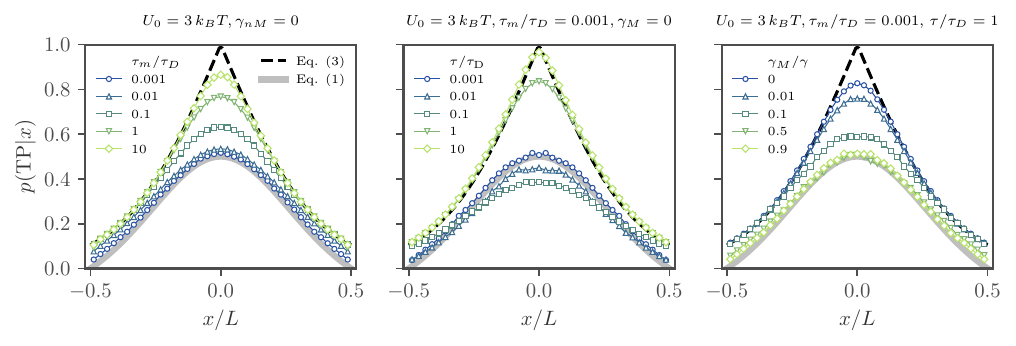}
  \put (6,34) {\LARGE \bf{A}}
  \put (37,34) {\LARGE \bf{B}}
  \put (68,34) {\LARGE \bf{C}}  
\end{overpic}
\caption{Transition-path probability profiles, $\ptpx$, between $x/L = \pm 0.5$ obtained from simulations of the \ac{GLE}, \cref{eq:gle} in a double-well potential. Analytical results are given in the overdamped limit by \cref{eq:ptp2} (gray solid lines) and in the inertial limit by \cref{eq:ptp2_inertial} (black broken lines). \textbf{A:} Results in the Markovian limit, $\gamma_{nM} = 0$, for various inertial time scales $\tm/\td$. \textbf{B:} Results in the overdamped non-Markovian limit,  $\gamma_M = 0$ and $\tm/\td = 0.001$, for various memory time scales $\tg/\td$. \textbf{C:} Results in the overdamped limit, $\tm/\td=0.001$ and $\tau/\td = 1$, for various ratios of Markovian friction to total friction, $\gamma_M/\gamma$.}
\label{fig:ptpxProfs}
\end{figure*}

Next, we give an estimate for the maximal value of $\ptpx$ using the transmission coefficient $\kappa$ predicted by rate theory, including inertia and non-Markovian friction, (see \ac{SM} \ref{SIptpx_nonMkv} for a detailed derivation)
\begin{align}
  \ptpxMax \approx \frac{(\kappa-1)^2+1}{(\kappa-2)^2},
  \label{eq:ptpMax_kappa}
\end{align}
with the properties $\ptpxMax=1$ for $\kappa=1$ and  $\ptpxMax=1/2$ for $\kappa=0$.
We employ the \ac{GH} prediction for the transmission coefficient \cite{Grote1980}, which accurately describes dynamics on the barrier \cite{Brunig2022c} 
\begin{align}
  \label{eq:kappaGH}
  \kappa = \frac{\lambda}{\OmegaMax},
\end{align}
where $\lambda$ is the \ac{GH} reactive frequency, which depends on inertial and non-Markovian effects as well as on the barrier shape (see \ac{SM} \ref{SIkappaGH}).
The frequency $\OmegaMax = \sqrt{|\UPPMax|/m}$ depends on the potential curvature at the barrier top $\UPPMax$. 
In the overdamped limit, $m \to 0$, we find
\begin{align}
  \label{eq:kappaGH_od}
  \kappa^2 = 1- \frac{\gamma}{\tau \UPPMax}.
\end{align}
Thus, combining \cref{eq:ptpMax_kappa,eq:kappaGH_od} we predict $\ptpxMax \to 1$ for $\tau \to \infty$.
Likewise, in \ac{SM} \ref{SIkappaGH}, we show that $\ptpxMax \to 1$ in the inertial non-Markovian limit, identical to the inertial Markovian limit predicted by \cref{eq:ptp2_inertial}.
Both \cref{eq:ptp2_inertial} and \cref{eq:ptpMax_kappa,eq:kappaGH} will be favorably compared with our simulations below.

\textbf{Simulation results:}
First, the Markovian limit is considered, given by $\gamma_{nM} \to 0$. 
Results for $\ptpx$ between $x = \pm L/2$ 
are shown in \cref{fig:ptpxProfs}A as colored markers for various values of the rescaled inertial time $\tm/\td$, which interpolate nicely between the analytical overdamped \cref{eq:ptp2} (gray solid line) and inertial limits \cref{eq:ptp2_inertial} (black broken line). 
The profiles differ most at their maxima at $x=0$, the transition state, for which values of $1/2$ in the overdamped limit and 1 in the inertial limit are predicted analytically.

\begin{figure*}[tbp]
  \centering
  \begin{overpic}[width=\textwidth]{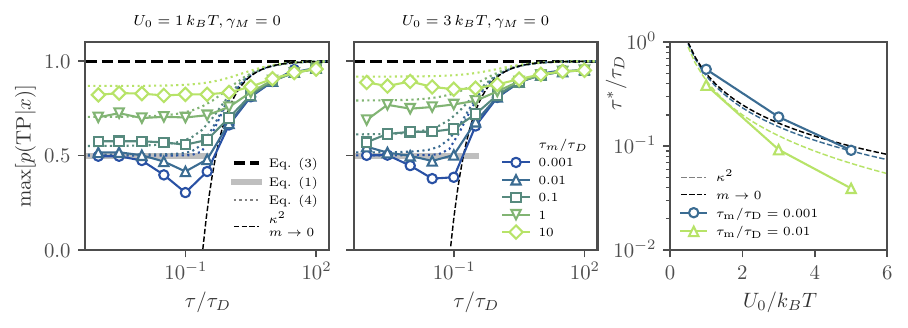}
  \put (6,34) {\LARGE \bf{A}}
  \put (37,34) {\LARGE \bf{B}}
  \put (68,34) {\LARGE \bf{C}}  
\end{overpic}
\vspace{-6mm}
\caption{\textbf{A, B:} Maximal values of the transition-path probability profiles, $\ptpxMax$, obtained from simulations of the \ac{GLE}, \cref{eq:gle} for $\gamma_M= 0$, plotted over the memory time scale $\tg/\td$ and shown for various inertial time scales $\tm / \td$ as different colored markers. 
The data is given for different barrier heights $U_0/k_BT=1.0$ (A) and $U_0/k_BT=3.0$ (B).
The analytical result in the overdamped limit, \cref{eq:ptp2}, is indicated as a gray horizontal line and in the inertial limit, \cref{eq:ptp2_inertial}, as a black broken horizontal line. A reference in the non-Markovian limit is given by \cref{eq:ptpMax_kappa}, employing \ac{GH} theory for the transmission coefficient $\kappa$, \cref{eq:kappaGH} (dotted lines for the different inertial time scales). The overdamped limit solution for $\kappa^2$, \cref{eq:kappaGH_od}, is given as a thin black broken line.
\textbf{C:} Crossover time $\ttu/\td$, defined as the memory time for which $\ptpx = 1/2$ is crossed in the long-memory limit, plotted in dependence of the barrier height $U_0$ and for two values of $\tm/\td$. Numerical data is shown as colored markers and the analytical estimates, employing \ac{GH} theory for transmission coefficient $\kappa$, \cref{eq:kappaGH}, as a broken lines. The overdamped limit solution for $\kappa^2$, \cref{eq:kappaGH_od}, is again given as a thin black broken line.
}
\label{fig:ptpxMaxs}
\end{figure*}

In \cref{fig:ptpxProfs}B, results for overdamped non-Markovian dynamics, $\gamma_M/\gamma = 0$ and $\tm/\td=0.001$, are shown as colored markers for various memory times $\tg/\td$.
For $\tau/\td \ll 1$ and $\tau/\td \gg 1$ the results asymptotically approach the Markovian and inertial limits, respectively, while for intermediate values of $\tau/\td$ 
(upward triangle and square markers) 
the maximal value of $\ptpx$ located at $x=0$ is below the analytical overdamped result of 1/2.
This is in contrast to results in the Markovian limit in \cref{fig:ptpxProfs}A, where $\ptpxMax \geq 1/2$ for all values of $\tm/\td$.
Importantly, the value $1/2$ that is indicative of a perfect reaction coordinate for overdamped Markovian dynamics  is obtained twice, not only for $\tg/\td \to 0$ but also for some intermediate value, which we call $\ttu/\td$ and analyze further in fig.~\ref{fig:ptpxMaxs}.

As is evident from \cref{fig:ptpxProfs}, the profiles differ most at there maximal value, $\ptpxMax$, which is analyzed in detail in the following.
In \cref{fig:ptpxMaxs}A and B, $\ptpxMax$ for non-Markovian dynamics, $\gamma_M=0$, is plotted over $\tg/\td$ for different barrier heights and various $\tm/\td$. 
For large $\tm/\td$ in the inertial limit the $\ptpxMax$ is large and approaches asymptotically the analytical limit of 1. 
Away from this limit, in the overdamped scenario for $\tm/\td \lesssim 0.01$, $\ptpxMax$ varies non-monotonically in dependence of the non-Markovian time scale $\tg/\td$.
For $\tg/\td \to 0$ and $\tm/\td \to 0$, in the overdamped Markovian limit, the analytical result $\ptpxMax = 1/2$ is obtained.
For increasing $\tg/\td$ a minimum with $\ptpxMax < 1/2$ appears and the limit  $\ptpxMax = 1$ is approached as $\tau/\td \to \infty$.  
While in the short-memory regime, $\tg/\td < \ttu/\td$, the value of $\ptpxMax$ depends strongly on both $\tm/\td$ and $\tg/\td$, in the long-memory regime, $\tg/\td > \ttu/\td$, non-Markovian effects are dominant and the results are independent of $\tm/\td$.

The analytical estimate \cref{eq:ptpMax_kappa} (using the \ac{GH} result \cref{eq:kappaGH} for $\kappa$) is given in \cref{fig:ptpxProfs}A and B as colored dotted lines, and agrees well with the numerical data away from the overdamped limit, i.e. for $\tm/\td\geq0.1$. Additionally, whenever $\kappa \geq 0.5$, i.e. in the inertial and non-Markovian limits, the analytical estimate reduces to the scaling of the squared transmission coefficient $\kappa^2$ by arguments given in \ac{SM} \ref{SIptpx_nonMkv}. 
This is supported by the favorable comparison of the analytical solution for $\kappa^2$ in the overdamped limit \cref{eq:kappaGH_od}, plotted in \cref{fig:ptpxMaxs} as thin back broken lines, to the numerical data for long memory times.
In fact, $\kappa^2$ is useful for analyzing the scaling of $\ttu$ in the following.

To quantify the non-monotonic scaling of $\ptpxMax$ in detail, in \cref{fig:ptpxMaxs}C, we show $\ttu/\td$, defined as the point at which $\ptpx = 1/2$ is crossed by piece-wise linear interpolation of the numerical data in dependence of $\tg/\td$, for two rather small values of $\tm/\td$ as relevant for protein folding. $\ttu/\td$ 
shifts to smaller values with increasing barrier height $U_0$, as well as with increasing $\tm/\td$. Importantly, the values for $\ttu/\td$ are in the range of memory times of fast-folding proteins  \cite{Dalton2023}, such as $\alpha$3D in \cref{ptpr_intro} for which $\tau/\td=0.2$ (defined as the first moment of the friction kernel).
Note, that estimates using the condition $\kappa^2=0.5 \approx \ptpxMax$ (broken lines) also satisfactorily explain the shifting of $\ttu/\td$.

At last, the special case of mixed Markovian and non-Markovian friction is considered by choosing $0 < \gamma_M/\gamma < 1$. 
Previously, $\ptpxMax \leq 0.5$ was found for this scenario using an overdamped formulation of the \ac{GLE} \cite{Berezhkovskii2018}, which is in contrast to the results for $\gamma_M=0$  in \cref{fig:ptpxMaxs} showing $\ptpxMax \to 1$ in the limit $\tau/\td \to \infty$. 
To elucidate on this, in \cref{fig:ptpxProfs}C, profiles of $\ptpx$ are given in the overdamped non-Markovian limit ($\tm/\td = 0.001$ and $\tau/\td =1 $) with gradually added Markovian friction, i.e. increasing $\gamma_M/\gamma$.
It transpires that already for a small ratio of Markovian friction to total friction, $\gamma_M/\gamma = 0.01$, $\ptpxMax$ is affected compared to the value for $\gamma_M = 0$, and decreases with further increasing $\gamma_M/\gamma$ ($\gamma_M/\gamma = 0.02$ is the smallest ratio considered in \cite{Berezhkovskii2018}).
 A comparison of this numerical data to the estimate \cref{eq:ptpMax_kappa} is presented in \ac{SM} \ref{SImixedResults} and analytical arguments that explain the decrease of $\ptpxMax$ with respect to $\gamma_M$ based on the asymptotic scaling of the transmission coefficient $\kappa$ are given in \ac{SM} \ref{SIkappaGH}.
The analysis reconciles the present study with previous results that indicated a strict decrease of $\ptpxMax$ in the non-Markovian limit \cite{Berezhkovskii2018}, and shows the subtle but significant effects arising for mixed Markovian and non-Markovian friction.

We analyzed the transition-path probability $\ptpx$ in a model double-well potential under the influence of non-Markovian memory friction and inertial effects.
Our results show that the maximal value of $1/2$ in the overdamped Markovian limit, which has been used as a benchmark for quantifying reaction-coordinate quality, is exceeded both in the presence of non-Markovian friction as well as inertial effects and a maximal value of $\ptpxMax=1$ is approached in either limit.
It is noteworthy that these results deviate from previous reports which suggested that for non-Markovian dynamics the maximal value of $\ptpx$ is always below the overdamped limit $\ptpxMax < 1/2$ \cite{Berezhkovskii2018}.
We show that the deviations result from subtle crossover effects of mixed Markovian and non-Markovian friction. 

We present an analytical estimate for $\ptpxMax$ based on the transmission coefficient, and find good agreement with the numerical data except in the overdamped limit for intermediate memory time.
Further work will need to investigate effects of competing different memory friction time scales.

\subsection{Acknowledgments}
We gratefully acknowledge support by the Deutsche Forschungsgemeinschaft (DFG) grant SFB 1078, by the European Research Council under the Horizon 2020 Programme, ERC Grant Agreement No. 835117, and computing time on the HPC clusters at the Physics Department. We thank the Shaw group for providing the simulation data of the $\alpha$3D protein.

\subsection*{Author contributions}
F.N.B. and R.R.N. conceived the theory and designed the simulations. F.N.B. performed simulations. F.N.B. analyzed the data. All authors discussed the results, analyses, and interpretations. F.N.B. and R.R.N. wrote the paper with input from all authors. The authors declare no competing interests.

\nocite{Hanggi1990,Kramers1940,Smoluchowski1906,Weiss1967,Zwanzig2001,Onsager1938}

\bibliography{bib}

\begin{thebibliography}{49}%
\makeatletter
\providecommand \@ifxundefined [1]{%
 \@ifx{#1\undefined}
}%
\providecommand \@ifnum [1]{%
 \ifnum #1\expandafter \@firstoftwo
 \else \expandafter \@secondoftwo
 \fi
}%
\providecommand \@ifx [1]{%
 \ifx #1\expandafter \@firstoftwo
 \else \expandafter \@secondoftwo
 \fi
}%
\providecommand \natexlab [1]{#1}%
\providecommand \enquote  [1]{``#1''}%
\providecommand \bibnamefont  [1]{#1}%
\providecommand \bibfnamefont [1]{#1}%
\providecommand \citenamefont [1]{#1}%
\providecommand \href@noop [0]{\@secondoftwo}%
\providecommand \href [0]{\begingroup \@sanitize@url \@href}%
\providecommand \@href[1]{\@@startlink{#1}\@@href}%
\providecommand \@@href[1]{\endgroup#1\@@endlink}%
\providecommand \@sanitize@url [0]{\catcode `\\12\catcode `\$12\catcode `\&12\catcode `\#12\catcode `\^12\catcode `\_12\catcode `\%12\relax}%
\providecommand \@@startlink[1]{}%
\providecommand \@@endlink[0]{}%
\providecommand \url  [0]{\begingroup\@sanitize@url \@url }%
\providecommand \@url [1]{\endgroup\@href {#1}{\urlprefix }}%
\providecommand \urlprefix  [0]{URL }%
\providecommand \Eprint [0]{\href }%
\providecommand \doibase [0]{https://doi.org/}%
\providecommand \selectlanguage [0]{\@gobble}%
\providecommand \bibinfo  [0]{\@secondoftwo}%
\providecommand \bibfield  [0]{\@secondoftwo}%
\providecommand \translation [1]{[#1]}%
\providecommand \BibitemOpen [0]{}%
\providecommand \bibitemStop [0]{}%
\providecommand \bibitemNoStop [0]{.\EOS\space}%
\providecommand \EOS [0]{\spacefactor3000\relax}%
\providecommand \BibitemShut  [1]{\csname bibitem#1\endcsname}%
\let\auto@bib@innerbib\@empty
\bibitem [{\citenamefont {Zijlstra}\ \emph {et~al.}(2020)\citenamefont {Zijlstra}, \citenamefont {Nettels}, \citenamefont {Satija}, \citenamefont {Makarov},\ and\ \citenamefont {Schuler}}]{Zijlstra2020}%
  \BibitemOpen
  \bibfield  {author} {\bibinfo {author} {\bibfnamefont {N.}~\bibnamefont {Zijlstra}}, \bibinfo {author} {\bibfnamefont {D.}~\bibnamefont {Nettels}}, \bibinfo {author} {\bibfnamefont {R.}~\bibnamefont {Satija}}, \bibinfo {author} {\bibfnamefont {D.~E.}\ \bibnamefont {Makarov}},\ and\ \bibinfo {author} {\bibfnamefont {B.}~\bibnamefont {Schuler}},\ }\bibfield  {title} {\bibinfo {title} {Transition {{Path Dynamics}} of a {{Dielectric Particle}} in a {{Bistable Optical Trap}}},\ }\href {https://doi.org/10.1103/PhysRevLett.125.146001} {\bibfield  {journal} {\bibinfo  {journal} {Phys. Rev. Lett.}\ }\textbf {\bibinfo {volume} {125}},\ \bibinfo {pages} {146001} (\bibinfo {year} {2020})}\BibitemShut {NoStop}%
\bibitem [{\citenamefont {Dudko}\ \emph {et~al.}(2011)\citenamefont {Dudko}, \citenamefont {Graham},\ and\ \citenamefont {Best}}]{Dudko2011}%
  \BibitemOpen
  \bibfield  {author} {\bibinfo {author} {\bibfnamefont {O.~K.}\ \bibnamefont {Dudko}}, \bibinfo {author} {\bibfnamefont {T.~G.~W.}\ \bibnamefont {Graham}},\ and\ \bibinfo {author} {\bibfnamefont {R.~B.}\ \bibnamefont {Best}},\ }\bibfield  {title} {\bibinfo {title} {Locating the {{Barrier}} for {{Folding}} of {{Single Molecules}} under an {{External Force}}},\ }\href {https://doi.org/10.1103/PhysRevLett.107.208301} {\bibfield  {journal} {\bibinfo  {journal} {Phys. Rev. Lett.}\ }\textbf {\bibinfo {volume} {107}},\ \bibinfo {pages} {208301} (\bibinfo {year} {2011})}\BibitemShut {NoStop}%
\bibitem [{\citenamefont {Truex}\ \emph {et~al.}(2015)\citenamefont {Truex}, \citenamefont {Chung}, \citenamefont {Louis},\ and\ \citenamefont {Eaton}}]{Truex2015}%
  \BibitemOpen
  \bibfield  {author} {\bibinfo {author} {\bibfnamefont {K.}~\bibnamefont {Truex}}, \bibinfo {author} {\bibfnamefont {H.~S.}\ \bibnamefont {Chung}}, \bibinfo {author} {\bibfnamefont {J.~M.}\ \bibnamefont {Louis}},\ and\ \bibinfo {author} {\bibfnamefont {W.~A.}\ \bibnamefont {Eaton}},\ }\bibfield  {title} {\bibinfo {title} {Testing {{Landscape Theory}} for {{Biomolecular Processes}} with {{Single Molecule Fluorescence Spectroscopy}}},\ }\href {https://doi.org/10.1103/PhysRevLett.115.018101} {\bibfield  {journal} {\bibinfo  {journal} {Phys. Rev. Lett.}\ }\textbf {\bibinfo {volume} {115}},\ \bibinfo {pages} {018101} (\bibinfo {year} {2015})}\BibitemShut {NoStop}%
\bibitem [{\citenamefont {Neupane}\ \emph {et~al.}(2012)\citenamefont {Neupane}, \citenamefont {Ritchie}, \citenamefont {Yu}, \citenamefont {Foster}, \citenamefont {Wang},\ and\ \citenamefont {Woodside}}]{Neupane2012}%
  \BibitemOpen
  \bibfield  {author} {\bibinfo {author} {\bibfnamefont {K.}~\bibnamefont {Neupane}}, \bibinfo {author} {\bibfnamefont {D.~B.}\ \bibnamefont {Ritchie}}, \bibinfo {author} {\bibfnamefont {H.}~\bibnamefont {Yu}}, \bibinfo {author} {\bibfnamefont {D.~A.~N.}\ \bibnamefont {Foster}}, \bibinfo {author} {\bibfnamefont {F.}~\bibnamefont {Wang}},\ and\ \bibinfo {author} {\bibfnamefont {M.~T.}\ \bibnamefont {Woodside}},\ }\bibfield  {title} {\bibinfo {title} {Transition {{Path Times}} for {{Nucleic Acid Folding Determined}} from {{Energy-Landscape Analysis}} of {{Single-Molecule Trajectories}}},\ }\href {https://doi.org/10.1103/PhysRevLett.109.068102} {\bibfield  {journal} {\bibinfo  {journal} {Phys. Rev. Lett.}\ }\textbf {\bibinfo {volume} {109}},\ \bibinfo {pages} {068102} (\bibinfo {year} {2012})}\BibitemShut {NoStop}%
\bibitem [{\citenamefont {Frederickx}\ \emph {et~al.}(2014)\citenamefont {Frederickx}, \citenamefont {{in't Veld}},\ and\ \citenamefont {Carlon}}]{Frederickx2014}%
  \BibitemOpen
  \bibfield  {author} {\bibinfo {author} {\bibfnamefont {R.}~\bibnamefont {Frederickx}}, \bibinfo {author} {\bibfnamefont {T.}~\bibnamefont {{in't Veld}}},\ and\ \bibinfo {author} {\bibfnamefont {E.}~\bibnamefont {Carlon}},\ }\bibfield  {title} {\bibinfo {title} {Anomalous {{Dynamics}} of {{DNA Hairpin Folding}}},\ }\href {https://doi.org/10.1103/PhysRevLett.112.198102} {\bibfield  {journal} {\bibinfo  {journal} {Phys. Rev. Lett.}\ }\textbf {\bibinfo {volume} {112}},\ \bibinfo {pages} {198102} (\bibinfo {year} {2014})}\BibitemShut {NoStop}%
\bibitem [{\citenamefont {Peters}(2016)}]{Peters2016}%
  \BibitemOpen
  \bibfield  {author} {\bibinfo {author} {\bibfnamefont {B.}~\bibnamefont {Peters}},\ }\bibfield  {title} {\bibinfo {title} {Reaction {{Coordinates}} and {{Mechanistic Hypothesis Tests}}},\ }\href {https://doi.org/10.1146/annurev-physchem-040215-112215} {\bibfield  {journal} {\bibinfo  {journal} {Annu. Rev. Phys. Chem.}\ }\textbf {\bibinfo {volume} {67}},\ \bibinfo {pages} {669} (\bibinfo {year} {2016})}\BibitemShut {NoStop}%
\bibitem [{\citenamefont {Mehdi}\ \emph {et~al.}(2022)\citenamefont {Mehdi}, \citenamefont {Wang}, \citenamefont {Pant},\ and\ \citenamefont {Tiwary}}]{Mehdi2022}%
  \BibitemOpen
  \bibfield  {author} {\bibinfo {author} {\bibfnamefont {S.}~\bibnamefont {Mehdi}}, \bibinfo {author} {\bibfnamefont {D.}~\bibnamefont {Wang}}, \bibinfo {author} {\bibfnamefont {S.}~\bibnamefont {Pant}},\ and\ \bibinfo {author} {\bibfnamefont {P.}~\bibnamefont {Tiwary}},\ }\bibfield  {title} {\bibinfo {title} {Accelerating {{All-Atom Simulations}} and {{Gaining Mechanistic Understanding}} of {{Biophysical Systems}} through {{State Predictive Information Bottleneck}}},\ }\href {https://doi.org/10.1021/acs.jctc.2c00058} {\bibfield  {journal} {\bibinfo  {journal} {J. Chem. Theory Comput.}\ }\textbf {\bibinfo {volume} {18}},\ \bibinfo {pages} {3231} (\bibinfo {year} {2022})}\BibitemShut {NoStop}%
\bibitem [{\citenamefont {Jung}\ \emph {et~al.}(2019)\citenamefont {Jung}, \citenamefont {Covino},\ and\ \citenamefont {Hummer}}]{Jung2019}%
  \BibitemOpen
  \bibfield  {author} {\bibinfo {author} {\bibfnamefont {H.}~\bibnamefont {Jung}}, \bibinfo {author} {\bibfnamefont {R.}~\bibnamefont {Covino}},\ and\ \bibinfo {author} {\bibfnamefont {G.}~\bibnamefont {Hummer}},\ }\href {https://doi.org/10.48550/arXiv.1901.04595} {\bibinfo {title} {Artificial {{Intelligence Assists Discovery}} of {{Reaction Coordinates}} and {{Mechanisms}} from {{Molecular Dynamics Simulations}}}} (\bibinfo {year} {2019}),\ \Eprint {https://arxiv.org/abs/1901.04595} {arXiv:1901.04595 [physics]} \BibitemShut {NoStop}%
\bibitem [{\citenamefont {Johnson}\ and\ \citenamefont {Hummer}(2012)}]{Johnson2012}%
  \BibitemOpen
  \bibfield  {author} {\bibinfo {author} {\bibfnamefont {M.~E.}\ \bibnamefont {Johnson}}\ and\ \bibinfo {author} {\bibfnamefont {G.}~\bibnamefont {Hummer}},\ }\bibfield  {title} {\bibinfo {title} {Characterization of a {{Dynamic String Method}} for the {{Construction}} of {{Transition Pathways}} in {{Molecular Reactions}}},\ }\href {https://doi.org/10.1021/jp212611k} {\bibfield  {journal} {\bibinfo  {journal} {J. Phys. Chem. B}\ }\textbf {\bibinfo {volume} {116}},\ \bibinfo {pages} {8573} (\bibinfo {year} {2012})}\BibitemShut {NoStop}%
\bibitem [{\citenamefont {Orland}(2011)}]{Orland2011}%
  \BibitemOpen
  \bibfield  {author} {\bibinfo {author} {\bibfnamefont {H.}~\bibnamefont {Orland}},\ }\bibfield  {title} {\bibinfo {title} {Generating transition paths by {{Langevin}} bridges},\ }\href {https://doi.org/10.1063/1.3586036} {\bibfield  {journal} {\bibinfo  {journal} {J. Chem. Phys.}\ }\textbf {\bibinfo {volume} {134}},\ \bibinfo {pages} {174114} (\bibinfo {year} {2011})}\BibitemShut {NoStop}%
\bibitem [{\citenamefont {Kim}\ and\ \citenamefont {Netz}(2015)}]{Kim2015}%
  \BibitemOpen
  \bibfield  {author} {\bibinfo {author} {\bibfnamefont {W.~K.}\ \bibnamefont {Kim}}\ and\ \bibinfo {author} {\bibfnamefont {R.~R.}\ \bibnamefont {Netz}},\ }\bibfield  {title} {\bibinfo {title} {The mean shape of transition and first-passage paths},\ }\href {https://doi.org/10.1063/1.4936408} {\bibfield  {journal} {\bibinfo  {journal} {J. Chem. Phys.}\ }\textbf {\bibinfo {volume} {143}},\ \bibinfo {pages} {224108} (\bibinfo {year} {2015})}\BibitemShut {NoStop}%
\bibitem [{\citenamefont {Daldrop}\ \emph {et~al.}(2016)\citenamefont {Daldrop}, \citenamefont {Kim},\ and\ \citenamefont {Netz}}]{Daldrop2016}%
  \BibitemOpen
  \bibfield  {author} {\bibinfo {author} {\bibfnamefont {J.~O.}\ \bibnamefont {Daldrop}}, \bibinfo {author} {\bibfnamefont {W.~K.}\ \bibnamefont {Kim}},\ and\ \bibinfo {author} {\bibfnamefont {R.~R.}\ \bibnamefont {Netz}},\ }\bibfield  {title} {\bibinfo {title} {Transition paths are hot},\ }\href {https://doi.org/10.1209/0295-5075/113/18004} {\bibfield  {journal} {\bibinfo  {journal} {Europhys. Lett.}\ }\textbf {\bibinfo {volume} {113}},\ \bibinfo {pages} {18004} (\bibinfo {year} {2016})}\BibitemShut {NoStop}%
\bibitem [{\citenamefont {Carlon}\ \emph {et~al.}(2018)\citenamefont {Carlon}, \citenamefont {Orland}, \citenamefont {Sakaue},\ and\ \citenamefont {Vanderzande}}]{Carlon2018}%
  \BibitemOpen
  \bibfield  {author} {\bibinfo {author} {\bibfnamefont {E.}~\bibnamefont {Carlon}}, \bibinfo {author} {\bibfnamefont {H.}~\bibnamefont {Orland}}, \bibinfo {author} {\bibfnamefont {T.}~\bibnamefont {Sakaue}},\ and\ \bibinfo {author} {\bibfnamefont {C.}~\bibnamefont {Vanderzande}},\ }\bibfield  {title} {\bibinfo {title} {Effect of {{Memory}} and {{Active Forces}} on {{Transition Path Time Distributions}}},\ }\href {https://doi.org/10.1021/acs.jpcb.8b06379} {\bibfield  {journal} {\bibinfo  {journal} {J. Phys. Chem. B}\ }\textbf {\bibinfo {volume} {122}},\ \bibinfo {pages} {11186} (\bibinfo {year} {2018})}\BibitemShut {NoStop}%
\bibitem [{\citenamefont {Louwerse}\ and\ \citenamefont {Sivak}(2022)}]{Louwerse2022}%
  \BibitemOpen
  \bibfield  {author} {\bibinfo {author} {\bibfnamefont {M.~D.}\ \bibnamefont {Louwerse}}\ and\ \bibinfo {author} {\bibfnamefont {D.~A.}\ \bibnamefont {Sivak}},\ }\bibfield  {title} {\bibinfo {title} {Information {{Thermodynamics}} of the {{Transition-Path Ensemble}}},\ }\href {https://doi.org/10.1103/PhysRevLett.128.170602} {\bibfield  {journal} {\bibinfo  {journal} {Phys. Rev. Lett.}\ }\textbf {\bibinfo {volume} {128}},\ \bibinfo {pages} {170602} (\bibinfo {year} {2022})}\BibitemShut {NoStop}%
\bibitem [{\citenamefont {Hummer}(2004)}]{Hummer2004}%
  \BibitemOpen
  \bibfield  {author} {\bibinfo {author} {\bibfnamefont {G.}~\bibnamefont {Hummer}},\ }\bibfield  {title} {\bibinfo {title} {From transition paths to transition states and rate coefficients},\ }\href {https://doi.org/10.1063/1.1630572} {\bibfield  {journal} {\bibinfo  {journal} {J. Chem. Phys.}\ }\textbf {\bibinfo {volume} {120}},\ \bibinfo {pages} {516} (\bibinfo {year} {2004})}\BibitemShut {NoStop}%
\bibitem [{\citenamefont {Dellago}\ \emph {et~al.}(1998)\citenamefont {Dellago}, \citenamefont {Bolhuis},\ and\ \citenamefont {Chandler}}]{Dellago1998}%
  \BibitemOpen
  \bibfield  {author} {\bibinfo {author} {\bibfnamefont {C.}~\bibnamefont {Dellago}}, \bibinfo {author} {\bibfnamefont {P.~G.}\ \bibnamefont {Bolhuis}},\ and\ \bibinfo {author} {\bibfnamefont {D.}~\bibnamefont {Chandler}},\ }\bibfield  {title} {\bibinfo {title} {Efficient transition path sampling: {{Application}} to {{Lennard-Jones}} cluster rearrangements},\ }\href {https://doi.org/10.1063/1.476378} {\bibfield  {journal} {\bibinfo  {journal} {J. Chem. Phys.}\ }\textbf {\bibinfo {volume} {108}},\ \bibinfo {pages} {9236} (\bibinfo {year} {1998})}\BibitemShut {NoStop}%
\bibitem [{\citenamefont {Bolhuis}\ \emph {et~al.}(2002)\citenamefont {Bolhuis}, \citenamefont {Chandler}, \citenamefont {Dellago},\ and\ \citenamefont {Geissler}}]{Bolhuis2002}%
  \BibitemOpen
  \bibfield  {author} {\bibinfo {author} {\bibfnamefont {P.~G.}\ \bibnamefont {Bolhuis}}, \bibinfo {author} {\bibfnamefont {D.}~\bibnamefont {Chandler}}, \bibinfo {author} {\bibfnamefont {C.}~\bibnamefont {Dellago}},\ and\ \bibinfo {author} {\bibfnamefont {P.~L.}\ \bibnamefont {Geissler}},\ }\bibfield  {title} {\bibinfo {title} {{{TRANSITION PATH SAMPLING}}: {{Throwing Ropes Over Rough Mountain Passes}}, in the {{Dark}}},\ }\href {https://doi.org/10.1146/annurev.physchem.53.082301.113146} {\bibfield  {journal} {\bibinfo  {journal} {Annu. Rev. Phys. Chem.}\ }\textbf {\bibinfo {volume} {53}},\ \bibinfo {pages} {291} (\bibinfo {year} {2002})}\BibitemShut {NoStop}%
\bibitem [{\citenamefont {Allen}\ \emph {et~al.}(2005)\citenamefont {Allen}, \citenamefont {Warren},\ and\ \citenamefont {{ten Wolde}}}]{Allen2005}%
  \BibitemOpen
  \bibfield  {author} {\bibinfo {author} {\bibfnamefont {R.~J.}\ \bibnamefont {Allen}}, \bibinfo {author} {\bibfnamefont {P.~B.}\ \bibnamefont {Warren}},\ and\ \bibinfo {author} {\bibfnamefont {P.~R.}\ \bibnamefont {{ten Wolde}}},\ }\bibfield  {title} {\bibinfo {title} {Sampling {{Rare Switching Events}} in {{Biochemical Networks}}},\ }\href {https://doi.org/10.1103/PhysRevLett.94.018104} {\bibfield  {journal} {\bibinfo  {journal} {Phys. Rev. Lett.}\ }\textbf {\bibinfo {volume} {94}},\ \bibinfo {pages} {018104} (\bibinfo {year} {2005})}\BibitemShut {NoStop}%
\bibitem [{\citenamefont {Sega}\ \emph {et~al.}(2007)\citenamefont {Sega}, \citenamefont {Faccioli}, \citenamefont {Pederiva}, \citenamefont {Garberoglio},\ and\ \citenamefont {Orland}}]{Sega2007}%
  \BibitemOpen
  \bibfield  {author} {\bibinfo {author} {\bibfnamefont {M.}~\bibnamefont {Sega}}, \bibinfo {author} {\bibfnamefont {P.}~\bibnamefont {Faccioli}}, \bibinfo {author} {\bibfnamefont {F.}~\bibnamefont {Pederiva}}, \bibinfo {author} {\bibfnamefont {G.}~\bibnamefont {Garberoglio}},\ and\ \bibinfo {author} {\bibfnamefont {H.}~\bibnamefont {Orland}},\ }\bibfield  {title} {\bibinfo {title} {Quantitative {{Protein Dynamics}} from {{Dominant Folding Pathways}}},\ }\href {https://doi.org/10.1103/PhysRevLett.99.118102} {\bibfield  {journal} {\bibinfo  {journal} {Phys. Rev. Lett.}\ }\textbf {\bibinfo {volume} {99}},\ \bibinfo {pages} {118102} (\bibinfo {year} {2007})}\BibitemShut {NoStop}%
\bibitem [{\citenamefont {Von~Hansen}\ \emph {et~al.}(2011)\citenamefont {Von~Hansen}, \citenamefont {Sedlmeier}, \citenamefont {Hinczewski},\ and\ \citenamefont {Netz}}]{Hansen2011}%
  \BibitemOpen
  \bibfield  {author} {\bibinfo {author} {\bibfnamefont {Y.}~\bibnamefont {Von~Hansen}}, \bibinfo {author} {\bibfnamefont {F.}~\bibnamefont {Sedlmeier}}, \bibinfo {author} {\bibfnamefont {M.}~\bibnamefont {Hinczewski}},\ and\ \bibinfo {author} {\bibfnamefont {R.~R.}\ \bibnamefont {Netz}},\ }\bibfield  {title} {\bibinfo {title} {Friction contribution to water-bond breakage kinetics},\ }\href {https://doi.org/10.1103/PhysRevE.84.051501} {\bibfield  {journal} {\bibinfo  {journal} {Phys. Rev. E}\ }\textbf {\bibinfo {volume} {84}},\ \bibinfo {pages} {051501} (\bibinfo {year} {2011})}\BibitemShut {NoStop}%
\bibitem [{\citenamefont {Hinczewski}\ \emph {et~al.}(2010)\citenamefont {Hinczewski}, \citenamefont {{von Hansen}}, \citenamefont {Dzubiella},\ and\ \citenamefont {Netz}}]{Hinczewski2010}%
  \BibitemOpen
  \bibfield  {author} {\bibinfo {author} {\bibfnamefont {M.}~\bibnamefont {Hinczewski}}, \bibinfo {author} {\bibfnamefont {Y.}~\bibnamefont {{von Hansen}}}, \bibinfo {author} {\bibfnamefont {J.}~\bibnamefont {Dzubiella}},\ and\ \bibinfo {author} {\bibfnamefont {R.~R.}\ \bibnamefont {Netz}},\ }\bibfield  {title} {\bibinfo {title} {How the diffusivity profile reduces the arbitrariness of protein folding free energies},\ }\href {https://doi.org/10.1063/1.3442716} {\bibfield  {journal} {\bibinfo  {journal} {J. Chem. Phys.}\ }\textbf {\bibinfo {volume} {132}},\ \bibinfo {pages} {245103} (\bibinfo {year} {2010})}\BibitemShut {NoStop}%
\bibitem [{\citenamefont {Neupane}\ \emph {et~al.}(2016)\citenamefont {Neupane}, \citenamefont {Manuel},\ and\ \citenamefont {Woodside}}]{Neupane2016}%
  \BibitemOpen
  \bibfield  {author} {\bibinfo {author} {\bibfnamefont {K.}~\bibnamefont {Neupane}}, \bibinfo {author} {\bibfnamefont {A.~P.}\ \bibnamefont {Manuel}},\ and\ \bibinfo {author} {\bibfnamefont {M.~T.}\ \bibnamefont {Woodside}},\ }\bibfield  {title} {\bibinfo {title} {Protein folding trajectories can be described quantitatively by one-dimensional diffusion over measured energy landscapes},\ }\href {https://doi.org/10.1038/nphys3677} {\bibfield  {journal} {\bibinfo  {journal} {Nat. Phys.}\ }\textbf {\bibinfo {volume} {12}},\ \bibinfo {pages} {700} (\bibinfo {year} {2016})}\BibitemShut {NoStop}%
\bibitem [{\citenamefont {Best}\ and\ \citenamefont {Hummer}(2005)}]{Best2005}%
  \BibitemOpen
  \bibfield  {author} {\bibinfo {author} {\bibfnamefont {R.~B.}\ \bibnamefont {Best}}\ and\ \bibinfo {author} {\bibfnamefont {G.}~\bibnamefont {Hummer}},\ }\bibfield  {title} {\bibinfo {title} {Reaction coordinates and rates from transition paths},\ }\href {https://doi.org/10.1073/pnas.0408098102} {\bibfield  {journal} {\bibinfo  {journal} {Proc. Natl. Acad. Sci. U. S. A.}\ }\textbf {\bibinfo {volume} {102}},\ \bibinfo {pages} {6732} (\bibinfo {year} {2005})}\BibitemShut {NoStop}%
\bibitem [{\citenamefont {Peters}\ and\ \citenamefont {Trout}(2006)}]{Peters2006}%
  \BibitemOpen
  \bibfield  {author} {\bibinfo {author} {\bibfnamefont {B.}~\bibnamefont {Peters}}\ and\ \bibinfo {author} {\bibfnamefont {B.~L.}\ \bibnamefont {Trout}},\ }\bibfield  {title} {\bibinfo {title} {Obtaining reaction coordinates by likelihood maximization},\ }\href {https://doi.org/10.1063/1.2234477} {\bibfield  {journal} {\bibinfo  {journal} {J. Chem. Phys.}\ }\textbf {\bibinfo {volume} {125}},\ \bibinfo {pages} {054108} (\bibinfo {year} {2006})}\BibitemShut {NoStop}%
\bibitem [{\citenamefont {Peters}(2010{\natexlab{a}})}]{Peters2010a}%
  \BibitemOpen
  \bibfield  {author} {\bibinfo {author} {\bibfnamefont {B.}~\bibnamefont {Peters}},\ }\bibfield  {title} {\bibinfo {title} {Recent advances in transition path sampling: Accurate reaction coordinates, likelihood maximisation and diffusive barrier-crossing dynamics},\ }\href {https://doi.org/10.1080/08927020903536382} {\bibfield  {journal} {\bibinfo  {journal} {Mol. Simul.}\ }\textbf {\bibinfo {volume} {36}},\ \bibinfo {pages} {1265} (\bibinfo {year} {2010}{\natexlab{a}})}\BibitemShut {NoStop}%
\bibitem [{\citenamefont {Peters}(2010{\natexlab{b}})}]{Peters2010b}%
  \BibitemOpen
  \bibfield  {author} {\bibinfo {author} {\bibfnamefont {B.}~\bibnamefont {Peters}},\ }\bibfield  {title} {\bibinfo {title} {P({{TP}}{\textbar}q) peak maximization: {{Necessary}} but not sufficient for reaction coordinate accuracy},\ }\href {https://doi.org/10.1016/j.cplett.2010.05.069} {\bibfield  {journal} {\bibinfo  {journal} {Chem. Phys. Lett.}\ }\textbf {\bibinfo {volume} {494}},\ \bibinfo {pages} {100} (\bibinfo {year} {2010}{\natexlab{b}})}\BibitemShut {NoStop}%
\bibitem [{\citenamefont {Lee}\ \emph {et~al.}(2019)\citenamefont {Lee}, \citenamefont {Ahn},\ and\ \citenamefont {Darve}}]{Lee2019}%
  \BibitemOpen
  \bibfield  {author} {\bibinfo {author} {\bibfnamefont {H.~S.}\ \bibnamefont {Lee}}, \bibinfo {author} {\bibfnamefont {S.~H.}\ \bibnamefont {Ahn}},\ and\ \bibinfo {author} {\bibfnamefont {E.~F.}\ \bibnamefont {Darve}},\ }\bibfield  {title} {\bibinfo {title} {The multi-dimensional generalized {{Langevin}} equation for conformational motion of proteins},\ }\href {https://doi.org/10.1063/1.5055573} {\bibfield  {journal} {\bibinfo  {journal} {J. Chem. Phys.}\ }\textbf {\bibinfo {volume} {150}},\ \bibinfo {pages} {174113} (\bibinfo {year} {2019})}\BibitemShut {NoStop}%
\bibitem [{\citenamefont {Ayaz}\ \emph {et~al.}(2021)\citenamefont {Ayaz}, \citenamefont {Tepper}, \citenamefont {Br{\"u}nig}, \citenamefont {Kappler}, \citenamefont {Daldrop},\ and\ \citenamefont {Netz}}]{Ayaz2021}%
  \BibitemOpen
  \bibfield  {author} {\bibinfo {author} {\bibfnamefont {C.}~\bibnamefont {Ayaz}}, \bibinfo {author} {\bibfnamefont {L.}~\bibnamefont {Tepper}}, \bibinfo {author} {\bibfnamefont {F.~N.}\ \bibnamefont {Br{\"u}nig}}, \bibinfo {author} {\bibfnamefont {J.}~\bibnamefont {Kappler}}, \bibinfo {author} {\bibfnamefont {J.~O.}\ \bibnamefont {Daldrop}},\ and\ \bibinfo {author} {\bibfnamefont {R.~R.}\ \bibnamefont {Netz}},\ }\bibfield  {title} {\bibinfo {title} {Non-{{Markovian}} modeling of protein folding},\ }\href {https://doi.org/10.1073/pnas.2023856118} {\bibfield  {journal} {\bibinfo  {journal} {Proc. Natl. Acad. Sci. U. S. A.}\ }\textbf {\bibinfo {volume} {118}},\ \bibinfo {pages} {e2023856118} (\bibinfo {year} {2021})}\BibitemShut {NoStop}%
\bibitem [{\citenamefont {Br{\"u}nig}\ \emph {et~al.}(2022{\natexlab{a}})\citenamefont {Br{\"u}nig}, \citenamefont {Daldrop},\ and\ \citenamefont {Netz}}]{Brunig2022e}%
  \BibitemOpen
  \bibfield  {author} {\bibinfo {author} {\bibfnamefont {F.~N.}\ \bibnamefont {Br{\"u}nig}}, \bibinfo {author} {\bibfnamefont {J.~O.}\ \bibnamefont {Daldrop}},\ and\ \bibinfo {author} {\bibfnamefont {R.~R.}\ \bibnamefont {Netz}},\ }\bibfield  {title} {\bibinfo {title} {Pair-{{Reaction Dynamics}} in {{Water}}: {{Competition}} of {{Memory}}, {{Potential Shape}}, and {{Inertial Effects}}},\ }\href {https://doi.org/10.1021/acs.jpcb.2c05923} {\bibfield  {journal} {\bibinfo  {journal} {J. Phys. Chem. B}\ }\textbf {\bibinfo {volume} {126}},\ \bibinfo {pages} {10295} (\bibinfo {year} {2022}{\natexlab{a}})}\BibitemShut {NoStop}%
\bibitem [{\citenamefont {Dalton}\ \emph {et~al.}(2023)\citenamefont {Dalton}, \citenamefont {Ayaz}, \citenamefont {Kiefer}, \citenamefont {Klimek}, \citenamefont {Tepper},\ and\ \citenamefont {Netz}}]{Dalton2023}%
  \BibitemOpen
  \bibfield  {author} {\bibinfo {author} {\bibfnamefont {B.~A.}\ \bibnamefont {Dalton}}, \bibinfo {author} {\bibfnamefont {C.}~\bibnamefont {Ayaz}}, \bibinfo {author} {\bibfnamefont {H.}~\bibnamefont {Kiefer}}, \bibinfo {author} {\bibfnamefont {A.}~\bibnamefont {Klimek}}, \bibinfo {author} {\bibfnamefont {L.}~\bibnamefont {Tepper}},\ and\ \bibinfo {author} {\bibfnamefont {R.~R.}\ \bibnamefont {Netz}},\ }\bibfield  {title} {\bibinfo {title} {Fast protein folding is governed by memory-dependent friction},\ }\href {https://doi.org/10.1073/pnas.2220068120} {\bibfield  {journal} {\bibinfo  {journal} {Proc. Natl. Acad. Sci.}\ }\textbf {\bibinfo {volume} {120}},\ \bibinfo {pages} {e2220068120} (\bibinfo {year} {2023})}\BibitemShut {NoStop}%
\bibitem [{\citenamefont {Dalton}\ and\ \citenamefont {Netz}(2024)}]{Dalton2024}%
  \BibitemOpen
  \bibfield  {author} {\bibinfo {author} {\bibfnamefont {B.~A.}\ \bibnamefont {Dalton}}\ and\ \bibinfo {author} {\bibfnamefont {R.~R.}\ \bibnamefont {Netz}},\ }\href {https://doi.org/10.48550/arXiv.2401.12027} {\bibinfo {title} {{{pH}} modulates friction memory effects in protein folding}} (\bibinfo {year} {2024}),\ \Eprint {https://arxiv.org/abs/2401.12027} {arXiv:2401.12027 [physics]} \BibitemShut {NoStop}%
\bibitem [{\citenamefont {Dalton}\ \emph {et~al.}(2024)\citenamefont {Dalton}, \citenamefont {Klimek}, \citenamefont {Kiefer}, \citenamefont {Br{\"u}nig}, \citenamefont {Colinet}, \citenamefont {Tepper}, \citenamefont {Abbasi},\ and\ \citenamefont {Netz}}]{Dalton2024a}%
  \BibitemOpen
  \bibfield  {author} {\bibinfo {author} {\bibfnamefont {B.~A.}\ \bibnamefont {Dalton}}, \bibinfo {author} {\bibfnamefont {A.}~\bibnamefont {Klimek}}, \bibinfo {author} {\bibfnamefont {H.}~\bibnamefont {Kiefer}}, \bibinfo {author} {\bibfnamefont {F.~N.}\ \bibnamefont {Br{\"u}nig}}, \bibinfo {author} {\bibfnamefont {H.}~\bibnamefont {Colinet}}, \bibinfo {author} {\bibfnamefont {L.}~\bibnamefont {Tepper}}, \bibinfo {author} {\bibfnamefont {A.}~\bibnamefont {Abbasi}},\ and\ \bibinfo {author} {\bibfnamefont {R.~R.}\ \bibnamefont {Netz}},\ }\href {https://doi.org/10.48550/arXiv.2410.22588} {\bibinfo {title} {Memory and {{Friction}}: {{From}} the {{Nanoscale}} to the {{Macroscale}}}} (\bibinfo {year} {2024}),\ \Eprint {https://arxiv.org/abs/2410.22588} {arXiv:2410.22588 [physics]} \BibitemShut {NoStop}%
\bibitem [{\citenamefont {Ayaz}\ \emph {et~al.}(2022)\citenamefont {Ayaz}, \citenamefont {Scalfi}, \citenamefont {Dalton},\ and\ \citenamefont {Netz}}]{Ayaz2022}%
  \BibitemOpen
  \bibfield  {author} {\bibinfo {author} {\bibfnamefont {C.}~\bibnamefont {Ayaz}}, \bibinfo {author} {\bibfnamefont {L.}~\bibnamefont {Scalfi}}, \bibinfo {author} {\bibfnamefont {B.~A.}\ \bibnamefont {Dalton}},\ and\ \bibinfo {author} {\bibfnamefont {R.~R.}\ \bibnamefont {Netz}},\ }\bibfield  {title} {\bibinfo {title} {Generalized {{Langevin}} equation with a nonlinear potential of mean force and nonlinear memory friction from a hybrid projection scheme},\ }\href@noop {} {\bibfield  {journal} {\bibinfo  {journal} {Phys. Rev. E}\ }\textbf {\bibinfo {volume} {105}},\ \bibinfo {pages} {054138} (\bibinfo {year} {2022})}\BibitemShut {NoStop}%
\bibitem [{\citenamefont {Ballard}\ and\ \citenamefont {Dellago}(2012)}]{Ballard2012}%
  \BibitemOpen
  \bibfield  {author} {\bibinfo {author} {\bibfnamefont {A.~J.}\ \bibnamefont {Ballard}}\ and\ \bibinfo {author} {\bibfnamefont {C.}~\bibnamefont {Dellago}},\ }\bibfield  {title} {\bibinfo {title} {Toward the {{Mechanism}} of {{Ionic Dissociation}} in {{Water}}},\ }\href {https://doi.org/10.1021/jp309300b} {\bibfield  {journal} {\bibinfo  {journal} {J. Phys. Chem. B}\ }\textbf {\bibinfo {volume} {116}},\ \bibinfo {pages} {13490} (\bibinfo {year} {2012})}\BibitemShut {NoStop}%
\bibitem [{\citenamefont {Berezhkovskii}\ and\ \citenamefont {Makarov}(2018)}]{Berezhkovskii2018}%
  \BibitemOpen
  \bibfield  {author} {\bibinfo {author} {\bibfnamefont {A.~M.}\ \bibnamefont {Berezhkovskii}}\ and\ \bibinfo {author} {\bibfnamefont {D.~E.}\ \bibnamefont {Makarov}},\ }\bibfield  {title} {\bibinfo {title} {Single-{{Molecule Test}} for {{Markovianity}} of the {{Dynamics}} along a {{Reaction Coordinate}}},\ }\href {https://doi.org/10.1021/acs.jpclett.8b00956} {\bibfield  {journal} {\bibinfo  {journal} {J. Phys. Chem. Lett.}\ }\textbf {\bibinfo {volume} {9}},\ \bibinfo {pages} {2190} (\bibinfo {year} {2018})}\BibitemShut {NoStop}%
\bibitem [{\citenamefont {Grote}\ and\ \citenamefont {Hynes}(1980)}]{Grote1980}%
  \BibitemOpen
  \bibfield  {author} {\bibinfo {author} {\bibfnamefont {R.~F.}\ \bibnamefont {Grote}}\ and\ \bibinfo {author} {\bibfnamefont {J.~T.}\ \bibnamefont {Hynes}},\ }\bibfield  {title} {\bibinfo {title} {The stable states picture of chemical reactions. {{II}}. {{Rate}} constants for condensed and gas phase reaction models},\ }\href {https://doi.org/10.1063/1.440485} {\bibfield  {journal} {\bibinfo  {journal} {J. Chem. Phys.}\ }\textbf {\bibinfo {volume} {73}},\ \bibinfo {pages} {2715} (\bibinfo {year} {1980})}\BibitemShut {NoStop}%
\bibitem [{\citenamefont {{Lindorff-Larsen}}\ \emph {et~al.}(2011)\citenamefont {{Lindorff-Larsen}}, \citenamefont {Piana}, \citenamefont {Dror},\ and\ \citenamefont {Shaw}}]{Lindorff-Larsen2011}%
  \BibitemOpen
  \bibfield  {author} {\bibinfo {author} {\bibfnamefont {K.}~\bibnamefont {{Lindorff-Larsen}}}, \bibinfo {author} {\bibfnamefont {S.}~\bibnamefont {Piana}}, \bibinfo {author} {\bibfnamefont {R.~O.}\ \bibnamefont {Dror}},\ and\ \bibinfo {author} {\bibfnamefont {D.~E.}\ \bibnamefont {Shaw}},\ }\bibfield  {title} {\bibinfo {title} {How {{Fast-Folding Proteins Fold}}},\ }\href {https://doi.org/10.1126/science.1208351} {\bibfield  {journal} {\bibinfo  {journal} {Science}\ }\textbf {\bibinfo {volume} {334}},\ \bibinfo {pages} {517} (\bibinfo {year} {2011})}\BibitemShut {NoStop}%
\bibitem [{sup()}]{suppMat}%
  \BibitemOpen
  \href@noop {} {\bibinfo {title} {See attached {{Supplemental Material}}.}}\BibitemShut {Stop}%
\bibitem [{\citenamefont {Chung}\ \emph {et~al.}(2015)\citenamefont {Chung}, \citenamefont {{Piana-Agostinetti}}, \citenamefont {Shaw},\ and\ \citenamefont {Eaton}}]{Chung2015a}%
  \BibitemOpen
  \bibfield  {author} {\bibinfo {author} {\bibfnamefont {H.~S.}\ \bibnamefont {Chung}}, \bibinfo {author} {\bibfnamefont {S.}~\bibnamefont {{Piana-Agostinetti}}}, \bibinfo {author} {\bibfnamefont {D.~E.}\ \bibnamefont {Shaw}},\ and\ \bibinfo {author} {\bibfnamefont {W.~A.}\ \bibnamefont {Eaton}},\ }\bibfield  {title} {\bibinfo {title} {Structural origin of slow diffusion in protein folding},\ }\href {https://doi.org/10.1126/science.aab1369} {\bibfield  {journal} {\bibinfo  {journal} {Science}\ }\textbf {\bibinfo {volume} {349}},\ \bibinfo {pages} {1504} (\bibinfo {year} {2015})}\BibitemShut {NoStop}%
\bibitem [{\citenamefont {Kappler}\ \emph {et~al.}(2018)\citenamefont {Kappler}, \citenamefont {Daldrop}, \citenamefont {Br{\"u}nig}, \citenamefont {Boehle},\ and\ \citenamefont {Netz}}]{Kappler2018}%
  \BibitemOpen
  \bibfield  {author} {\bibinfo {author} {\bibfnamefont {J.}~\bibnamefont {Kappler}}, \bibinfo {author} {\bibfnamefont {J.~O.}\ \bibnamefont {Daldrop}}, \bibinfo {author} {\bibfnamefont {F.~N.}\ \bibnamefont {Br{\"u}nig}}, \bibinfo {author} {\bibfnamefont {M.~D.}\ \bibnamefont {Boehle}},\ and\ \bibinfo {author} {\bibfnamefont {R.~R.}\ \bibnamefont {Netz}},\ }\bibfield  {title} {\bibinfo {title} {Memory-induced acceleration and slowdown of barrier crossing},\ }\href {https://doi.org/10.1063/1.4998239} {\bibfield  {journal} {\bibinfo  {journal} {J. Chem. Phys.}\ }\textbf {\bibinfo {volume} {148}},\ \bibinfo {pages} {014903} (\bibinfo {year} {2018})}\BibitemShut {NoStop}%
\bibitem [{\citenamefont {Mori}(1965)}]{Mori1965a}%
  \BibitemOpen
  \bibfield  {author} {\bibinfo {author} {\bibfnamefont {H.}~\bibnamefont {Mori}},\ }\bibfield  {title} {\bibinfo {title} {Transport, collective motion, and {{Brownian}} motion},\ }\href {https://doi.org/10.1143/PTP.33.423} {\bibfield  {journal} {\bibinfo  {journal} {Prog. Theor. Phys.}\ }\textbf {\bibinfo {volume} {33}},\ \bibinfo {pages} {423} (\bibinfo {year} {1965})}\BibitemShut {NoStop}%
\bibitem [{\citenamefont {Zwanzig}(1960)}]{Zwanzig1960}%
  \BibitemOpen
  \bibfield  {author} {\bibinfo {author} {\bibfnamefont {R.}~\bibnamefont {Zwanzig}},\ }\bibfield  {title} {\bibinfo {title} {Ensemble {{Method}} in the {{Theory}} of {{Irreversibility}}},\ }\href {https://doi.org/10.1063/1.1731409} {\bibfield  {journal} {\bibinfo  {journal} {J. Chem. Phys.}\ }\textbf {\bibinfo {volume} {33}},\ \bibinfo {pages} {1338} (\bibinfo {year} {1960})}\BibitemShut {NoStop}%
\bibitem [{\citenamefont {Br{\"u}nig}\ \emph {et~al.}(2022{\natexlab{b}})\citenamefont {Br{\"u}nig}, \citenamefont {Netz},\ and\ \citenamefont {Kappler}}]{Brunig2022c}%
  \BibitemOpen
  \bibfield  {author} {\bibinfo {author} {\bibfnamefont {F.~N.}\ \bibnamefont {Br{\"u}nig}}, \bibinfo {author} {\bibfnamefont {R.~R.}\ \bibnamefont {Netz}},\ and\ \bibinfo {author} {\bibfnamefont {J.}~\bibnamefont {Kappler}},\ }\bibfield  {title} {\bibinfo {title} {Barrier-crossing times for different non-{{Markovian}} friction in well and barrier: {{A}} numerical study},\ }\href {https://doi.org/10.1103/PhysRevE.106.044133} {\bibfield  {journal} {\bibinfo  {journal} {Phys. Rev. E}\ }\textbf {\bibinfo {volume} {106}},\ \bibinfo {pages} {44133} (\bibinfo {year} {2022}{\natexlab{b}})}\BibitemShut {NoStop}%
\bibitem [{\citenamefont {H{\"a}nggi}\ \emph {et~al.}(1990)\citenamefont {H{\"a}nggi}, \citenamefont {Talkner},\ and\ \citenamefont {Borkovec}}]{Hanggi1990}%
  \BibitemOpen
  \bibfield  {author} {\bibinfo {author} {\bibfnamefont {P.}~\bibnamefont {H{\"a}nggi}}, \bibinfo {author} {\bibfnamefont {P.}~\bibnamefont {Talkner}},\ and\ \bibinfo {author} {\bibfnamefont {M.}~\bibnamefont {Borkovec}},\ }\bibfield  {title} {\bibinfo {title} {Reaction-rate theory: {{Fifty}} years after {{Kramers}}},\ }\href {https://doi.org/10.1103/RevModPhys.62.251} {\bibfield  {journal} {\bibinfo  {journal} {Rev. Mod. Phys.}\ }\textbf {\bibinfo {volume} {62}},\ \bibinfo {pages} {251} (\bibinfo {year} {1990})}\BibitemShut {NoStop}%
\bibitem [{\citenamefont {Kramers}(1940)}]{Kramers1940}%
  \BibitemOpen
  \bibfield  {author} {\bibinfo {author} {\bibfnamefont {H.~A.}\ \bibnamefont {Kramers}},\ }\bibfield  {title} {\bibinfo {title} {Brownian motion in a field of force and the diffusion model of chemical reactions},\ }\href {https://doi.org/10.1016/S0031-8914(40)90098-2} {\bibfield  {journal} {\bibinfo  {journal} {Physica}\ }\textbf {\bibinfo {volume} {7}},\ \bibinfo {pages} {284} (\bibinfo {year} {1940})}\BibitemShut {NoStop}%
\bibitem [{\citenamefont {{von Smoluchowski}}(1906)}]{Smoluchowski1906}%
  \BibitemOpen
  \bibfield  {author} {\bibinfo {author} {\bibfnamefont {M.}~\bibnamefont {{von Smoluchowski}}},\ }\bibfield  {title} {\bibinfo {title} {Zur kinetischen {{Theorie}} der {{Brownschen Molekularbewegung}} und der {{Suspensionen}}},\ }\href@noop {} {\bibfield  {journal} {\bibinfo  {journal} {Ann. Phys.}\ }\textbf {\bibinfo {volume} {326}},\ \bibinfo {pages} {756} (\bibinfo {year} {1906})}\BibitemShut {NoStop}%
\bibitem [{\citenamefont {Weiss}(1967)}]{Weiss1967}%
  \BibitemOpen
  \bibfield  {author} {\bibinfo {author} {\bibfnamefont {G.~H.}\ \bibnamefont {Weiss}},\ }\bibfield  {title} {\bibinfo {title} {First {{Passage Time Problems}} in {{Chemical Physics}}},\ }\href {https://doi.org/10.1002/9780470140154.ch1} {\bibfield  {journal} {\bibinfo  {journal} {Adv. Chem. Phys.}\ }\textbf {\bibinfo {volume} {13}},\ \bibinfo {pages} {1} (\bibinfo {year} {1967})}\BibitemShut {NoStop}%
\bibitem [{\citenamefont {Zwanzig}(2001)}]{Zwanzig2001}%
  \BibitemOpen
  \bibfield  {author} {\bibinfo {author} {\bibfnamefont {R.}~\bibnamefont {Zwanzig}},\ }\href@noop {} {\emph {\bibinfo {title} {Nonequilibrium {{Statistical Mechanics}}}}}\ (\bibinfo  {publisher} {Oxford University Press},\ \bibinfo {year} {2001})\BibitemShut {NoStop}%
\bibitem [{\citenamefont {Onsager}(1938)}]{Onsager1938}%
  \BibitemOpen
  \bibfield  {author} {\bibinfo {author} {\bibfnamefont {L.}~\bibnamefont {Onsager}},\ }\bibfield  {title} {\bibinfo {title} {Initial {{Recombination}} of {{Ions}}},\ }\href {https://doi.org/10.1103/PhysRev.54.554} {\bibfield  {journal} {\bibinfo  {journal} {Phys. Rev.}\ }\textbf {\bibinfo {volume} {54}},\ \bibinfo {pages} {554} (\bibinfo {year} {1938})}\BibitemShut {NoStop}%
\end{thebibliography}%


\begin{thebibliography}{0}%
\makeatletter
\providecommand \@ifxundefined [1]{%
 \@ifx{#1\undefined}
}%
\providecommand \@ifnum [1]{%
 \ifnum #1\expandafter \@firstoftwo
 \else \expandafter \@secondoftwo
 \fi
}%
\providecommand \@ifx [1]{%
 \ifx #1\expandafter \@firstoftwo
 \else \expandafter \@secondoftwo
 \fi
}%
\providecommand \natexlab [1]{#1}%
\providecommand \enquote  [1]{``#1''}%
\providecommand \bibnamefont  [1]{#1}%
\providecommand \bibfnamefont [1]{#1}%
\providecommand \citenamefont [1]{#1}%
\providecommand \href@noop [0]{\@secondoftwo}%
\providecommand \href [0]{\begingroup \@sanitize@url \@href}%
\providecommand \@href[1]{\@@startlink{#1}\@@href}%
\providecommand \@@href[1]{\endgroup#1\@@endlink}%
\providecommand \@sanitize@url [0]{\catcode `\\12\catcode `\$12\catcode `\&12\catcode `\#12\catcode `\^12\catcode `\_12\catcode `\%12\relax}%
\providecommand \@@startlink[1]{}%
\providecommand \@@endlink[0]{}%
\providecommand \url  [0]{\begingroup\@sanitize@url \@url }%
\providecommand \@url [1]{\endgroup\@href {#1}{\urlprefix }}%
\providecommand \urlprefix  [0]{URL }%
\providecommand \Eprint [0]{\href }%
\providecommand \doibase [0]{https://doi.org/}%
\providecommand \selectlanguage [0]{\@gobble}%
\providecommand \bibinfo  [0]{\@secondoftwo}%
\providecommand \bibfield  [0]{\@secondoftwo}%
\providecommand \translation [1]{[#1]}%
\providecommand \BibitemOpen [0]{}%
\providecommand \bibitemStop [0]{}%
\providecommand \bibitemNoStop [0]{.\EOS\space}%
\providecommand \EOS [0]{\spacefactor3000\relax}%
\providecommand \BibitemShut  [1]{\csname bibitem#1\endcsname}%
\let\auto@bib@innerbib\@empty
\end{thebibliography}%

\end{document}


\title{Supplemental material for:\\Non-Markovianity increases transition path probability}

\author{Florian N. Br\"unig}
\affiliation{Department of Physics, Freie Universität Berlin, 14195 Berlin, Germany}

\author{Benjamin A. Dalton}
\affiliation{Department of Physics, Freie Universität Berlin, 14195 Berlin, Germany}

\author{Jan O. Daldrop}
\affiliation{Department of Physics, Freie Universität Berlin, 14195 Berlin, Germany}

\author{Roland R. Netz}
\affiliation{Department of Physics, Freie Universität Berlin, 14195 Berlin, Germany}

\date{\today}

\maketitle

\begin{acronym}[Bash]
   \acro{GH}{Grote and Hynes}
   \acro{GLE}{generalized Langevin equation}
   \acro{LE}{Langevin equation}
   \acro{MD}{molecular dynamics}
   \acro{MSD}{mean squared displacement}
   \acro{PTP}{$p(\text{TP}|x)$}
   \acro{PMF}{potential of mean force}
   \acro{RC}{reaction coordinate}
   \acro{SM}{Supplementary Material}
   \acro{TP}{transition path}
   \acro{TS}{transition state}
\end{acronym}  

\label{paper_ptpr_appendix}

\renewcommand{\thesubsection}{\Roman{subsection}}
\setcounter{secnumdepth}{2}

\subsection{Numerical estimation of $\ptpx$}
\label{SI_method}

For the simulation analysis all trajectory paths in the barrier region between $x=\pm L/2$ are assigned to the transition-path ensemble or not, depending on whether they return to the initial boundary or cross the barrier to the other one. The spatial distributions of these two ensembles, $p(x,\text{TP})$ and $p(x,\neg \text{TP})$, are calculated, which sum to the equilibrium distribution $p_{eq}(x)=p(x,\text{TP})+p(x, \neg \text{TP})$. 
The fraction is written and using the Bayesian equality $p(x|\text{TP})p(\text{TP})=p(x,\text{TP})=p(\text{TP}|x)p_{eq}(x)$ it follows
\begin{align}
  \label{eq:ptp_ex}
\frac{p(x,\text{TP})}{p(x,\text{TP})+p(x, \neg \text{TP})}&=\frac{p(x|\text{TP})p(\text{TP})}{p_{eq}(x)}=p(\text{TP}|x),
\end{align}
which is the expression used to extract $p(\text{TP}|x)$ from equilibrium simulations.

\subsection{$\ptpx$ in the overdamped Markovian limit}
\label{SIptpx_od}

A transition path at point $x$, connecting boundaries $A$ and $B$, can be either the joint trajectory forward in time to $A$ and backward in time to $B$, or the reverse. Expressed in terms of commitor probabilities the following relation is obtained \cite{Hummer2004}
\begin{align}
\label{eq:ptp1}
p(\text{TP}|x)=\phi_A(x)\overline{\phi_B(x)}+\overline{\phi_A(x)}\phi_B(x),
\end{align}
where the overline $\overline{\phi_{A,B}}$ denotes evaluation of time reversed trajectories. For a derivation of the commitor probabilities in the overdamped Markovian limit refer to section \ref{SIcommitor}. \cref{eq:ptp1} is at the heart of transition-path sampling by shooting \cite{Bolhuis2002}.

Since in the overdamped Markovian limit also $\phi_A(x)=\overline{\phi_A(x)}$ follows from time-reversal symmetry, \cref{eq:ptp1} simplifies to \cite{Hummer2004}
\begin{align}
\label{eqSI:ptp2}
p(\text{TP}|x)=2\phi_A(x)\phi_B(x)=2\phi_A(x)(1-\phi_A(x)).
\end{align}

\subsection{Commitor probabilities in the overdamped Markovian limit}
\label{SIcommitor}

The commitor probabilities $\phi_{A}(x)$ and $\phi_{B}(x)$ are the probabilities that a system in configuration $x$ between two absorbing boundaries $x_A<x<x_B$ will first cross either $x_A$ or $x_B$ with
\begin{align}
\phi_{A}(x)+\phi_{B}(x)=1
\end{align}
in the overdamped Markovian limit. An expression for the commitor in this limit is derived in the following. 

A one-dimensional system obeying overdamped Markovian dynamics is described by the Fokker Planck equation, or Smoluchowski equation, for the probability density $\Psi(x,t)$ of finding the system at position $x$ at time $t$ \cite{Smoluchowski1906,Weiss1967}
\begin{align}
\label{fokker}
\frac{\partial}{\partial t} \Psi(x,t) = \frac{\partial}{\partial x} D(x) e^{-\beta U(x)} \frac{\partial}{\partial x} \Psi(x,t) e^{\beta U(x)},
\end{align}
where $U(x)$ is the potential energy and $D(x)$ the position-dependent diffusion coefficient.
\cref{fokker} is solved by a Greens function $G(x,t|x_0)$ for arbitrary initial conditions $x_0$.
The survival probability $S(x_0,t)$ for the system to reside at time $t$ within the interval $x_A<x<x_B$ with absorbing boundaries is given by \cite{Zwanzig2001,Kim2015}
\begin{align}
S(x_0,t) = \int_{x_A}^{x_B} dx' G(x',t|x_0).
\end{align}
Its time derivative can be interpreted as the sum of fluxes at the boundaries $j(x_{A/B},t|x_0)$
\begin{align}
\frac{\partial}{\partial t} S(x_0,t) = j(x_A,t|x_0)-j(x_B,t|x_0),
\end{align}
with the moments of the fluxes given by
\begin{align}
\label{fluxMom}
j^{(n)}(x_{A/B}|x_0)=\int_0^{\infty}t^n\ j(x_{A/B},t|x_0).
\end{align}
The fluxes also satisfy the adjoint Fokker Planck equation and thus the following relation is obtained by partial integration of \cref{fluxMom} using $j(x_{A/B},t|x_0)=0$ for $t=0$ and $t \to \infty$
\begin{align}
-n j^{(n-1)}(x_{A/B}|x_0) = e^{\beta U(x)} \frac{\partial}{\partial x} D(x) e^{-\beta U(x)} \frac{\partial}{\partial x} j^{(n)}(x_{A/B}|x_0) .
\end{align}
Interpreting the commitor probability $\phi_{A}(x)$ and $\phi_{B}(x)$ as the mean flux, i.e. the zeroth moment, it follows that $\phi_{A/B}(x)$ is a stationary solution to the adjoint Fokker Planck equation \cite{Onsager1938}
\begin{align}
  0 = e^{\beta U(x)} \frac{\partial}{\partial x} D(x) e^{-\beta U(x)} \frac{\partial}{\partial x} \phi_{A/B}(x).
\end{align}
The equation can be solved for $\phi_{B}(x)$ using the boundary conditions $\phi_{B}(x_A)=0$, where $C$ is a constant obtained from the first integration
\begin{align}
D (x') e^{-\beta U(x')} \frac{\partial}{\partial x'} \phi_{B}(x') &= C \notag \\
\int_{x_A}^{x}dx'\frac{\partial}{\partial x'}\phi_{B}(x') &= C \int_{x_A}^{x}dx'\ \frac{e^{\beta U(x')}}{D(x')}\notag  \\
\phi_{B}(x)&= C \int_{x_A}^{x}dx'\ \frac{e^{\beta U(x')}}{D(x')},
\end{align}
and using $\phi_{B}(x_B)=1$, leading to
\begin{align}
C^{-1} &= \int_{x_A}^{x_B}dx'\ \frac{e^{U(x')}}{D(x')}.
\end{align}
The same derivation leads to the result for $\phi_{A}(x)$
\begin{align}
\label{commitorA}
\phi_{A}(x)= C \int_{x}^{x_B}dx'\ \frac{e^{\beta U(x')}}{D(x')}.
\end{align}

\subsection{Numerical solution of the generalized Langevin equation}
\label{SIgle}

The \ac{GLE}, Eq.~\eqref{eq:gle} in the main text, which is repeated here for convenience
\begin{align}
m\ddot x (t) = &-\gamma_{M} \dot x(t) -\int_{0}^{t}\Gamma(t')\dot x(t-t')dt' -  \nabla U[x(t)] + \eta(t),
    \label{eqSI:gle}
\end{align}
is for exponential memory friction $\Gamma(t)=(\gamma_{nM}/\tg)\exp{-t/\tg}$ equivalent to the following coupled system of Markovian Langevin equations
\begin{align}
\dot{x}(t) &= v(t) \\
m \dot{v}(t) &= - \nabla U[x(t)] - \gamma_{M} v(t) + a [y(t)-x(t)] + F_M(t) \label{eq:equation_of_motion_x} \\
\gamma_{nM} \dot{y}(t) &= a[x(t)-y(t)]+F_{nM}(t), \label{eq:equation_of_motion_y}
\end{align}
with $a = \gamma_{nM}/ \tg$. The random forces $F_M(t)$ and $F_{nM}(t)$ each have zero mean $\langle F_M(t) \rangle = \langle F_{nM}(t) \rangle = 0 $ and are correlated according to $F_{M,nM} = 2 k_BT \gamma_{n,nM} \delta(t-t')$.

Using dimensionless units introduced in the main text, the \ac{GLE} Eq.~\eqref{eqSI:gle} can be rewritten as
\begin{align}
 \begin{split}
	\label{eq:gleTD}
	\frac{\tm}{\td}\ddot{\tilde{x}}(\tDL) = &- \left(1-\frac{\gamma_{M}}{\gamma}\right) \frac{\td}{\tg} \int_{0}^{\tDL} \exp\left[-\frac{\td}{\tg}\tDL'\right] \dot{\tilde{x}}(\tDL-\tDL')\,d\tDL'\\
  &- \frac{\gamma_{M}}{\gamma} \dot{\tilde{x}}(\tDL) +\FDL\left[\xDL(\tDL)\right] + \noiseDL(\tDL),
 \end{split}
\end{align}
where $\xDL(\tDL) := x(\td \tDL)/L$, $\dot{\tilde{x}}(\tDL) = \td \dot{x}(\td \tDL)/L$, $\ddot{\tilde{x}}(\tDL) = \td^2 \ddot{x}(\td \tDL)/L$ are dimensionless position, velocity and acceleration, $\noiseDL(\tDL)$ is the dimensionless random force with autocorrelation
\begin{align}
\langle \noiseDL(\tDL) \noiseDL(\tDL')\rangle= 2 \frac{\gamma_{M}}{\gamma} \delta  (\tDL-\tDL') +\left(1-\frac{\gamma_{M}}{\gamma}\right) \frac{\td}{\tg}\exp\left[-\frac{\td}{\tg}\left|\tDL-\tDL'\right|\right].
\label{eq:LangevinRFAC_DiffusiveTimescale}
\end{align}
$\FDL[\xDL(\tDL)] := - (L/\kT) \nabla U\left[L \xDL(\tDL)\right]$ is the dimensionless deterministic force for the double-well potential $U(\xDL)=U_0/(k_BT)(\xDL^2-1)^2$ given by
\begin{equation}
	\label{eq:QuarticPotentialFDL}
	\FDL(\xDL) = - 4 \frac{U_0}{k_BT} \left(\xDL^2-1\right) \xDL,
\end{equation}
with the barrier height $U_0/k_BT$.

Note, that in case of $m=0$ and $\gamma_M, F_M(t) = 0$, the equation of motion for $x(t)$, derived from \cref{eq:equation_of_motion_x} and \cref{eq:equation_of_motion_y}, 
\begin{align}
  x(t) &= y(t) - \frac{1}{a} \nabla U[x(t)] \\
 \dot{x}(t) &= \dot y(t) - \frac{1}{a}  U''[x(t)] \dot x(t) \\
  &= - \frac{1}{\gamma_{nM}} \nabla U[x(t)] + \frac{1}{\gamma_{nM}}F_{nM}(t) - \frac{1}{a}  U''[x(t)] \dot x(t),
\end{align}
may become singular and numerically unstable at the barrier, i.e. for $U''(x)<0$ and small $a$, as demonstrated by the resulting overdamped equation of motion for $x(t)$:
\begin{align}
\dot{x}(t) = \left( - \frac{1}{\gamma_{nM}} \nabla U[x(t)] + \frac{1}{\gamma_{nM}}F_{nM}(t) \right) \left(1 + \frac{1}{a}  U''[x(t)]\right)^{-1}.
\end{align}

\subsection{$\ptpx$ in the non-Markovian limit}
\label{SIptpx_nonMkv}

The commitor is useful to derive an exact expression of $\ptpx$ in the overdamped Markovian limit, as shown above in section \ref{SIptpx_od}.
Here, the transmission coefficient $\kappa$ is considered in order to obtain the scaling of $\ptpxMax$ away from the overdamped Markovian limit.
The transmission coefficient $\kappa$ is the ratio of the actual barrier-crossing rate $k$ and the transition-state rate $k_{\rm TST}=\OmegaMin \exp(U_0/k_BT)$, where $U_0$ is the potential barrier height and $\OmegaMin = \sqrt{|\UPPMin|/m}$ depends on the potential curvature in the minimum $\UPPMin$ \cite{Hanggi1990}
\begin{align}
    \kappa = \frac{k}{k_{\rm TST}}.
\end{align}
$k_{\rm TST}$ is an upper bound for the actual rate. While for overdamped and Markovian dynamics $\kappa$ is generally very small and the exact estimation is an extensive field of research \cite{Hanggi1990,Kappler2018}, for underdamped or non-Markovian systems, $\kappa \to 1$ is obtained as the transition-state rate is approached.

Since the transition-state rate is the fastest possible rate, and corresponds to forming a transition path at the first attempt, $\kappa^{-1}$ quantifies the average number of attempts required per successful transition path. Or in other words, $\kappa$ quantifies the transition-path probability given an initial velocity towards a product state $X$. 
By these arguments we relate $\kappa$ to  the velocity-direction-dependent commitor $\phi_{X}(x=0,v_{\to X})$ at the barrier top by summing over all doubly unsuccessful events, each with probability $(1-\kappa)$,
\begin{align}
  \phi_{X}(0,v_{\to X}) &= \kappa \sum_{n=0}^{\infty} (1-\kappa)^{2n} = \frac{1}{2-\kappa}\\
  \phi_{X}(0,-v_{\to X}) &= \kappa \sum_{n=0}^{\infty} (1-\kappa)^{2n+1} = \frac{1-\kappa}{2-\kappa}.
\end{align}

We expand the transition-path probability in the velocity coordinate and split the velocity distribution in velocities pointing towards $A$ or $B$ respectively, to obtain
\begin{align}
  \ptpxMax &= \langle (p(\mathrm{TP}|x=0,v) \rangle_v \nonumber \\
  &= \frac{1}{2}\langle p(\mathrm{TP}|0,v_{\to A}) + p(\mathrm{TP}|0,v_{\to B}) \rangle_v.
\end{align}
Following arguments by \citet{Berezhkovskii2018}, $ \phi_B(0,-v_{\to A})=1-\phi_A(0,-v_{\to A}) =  \phi_A(0,v_{\to A})$, again based on symmetry, from which follows,
\begin{align}
  p(\mathrm{TP}|0,v_{\to A}) &= \phi_A(0,v_{\to A})\phi_B(0,-v_{\to A})\\
  &\ \ \ + \phi_A(0,-v_{\to A})\phi_B(0,v_{\to A}) \\
    &=  \phi^2_A(0,v_{\to A})+ \phi^2_A(0,-v_{\to A}).
\end{align}
We therefore arrive at
\begin{align}
  \ptpxMax \approx &\left( \frac{1}{\kappa-2} \right)^2 + \left( \frac{\kappa-1}{\kappa-2} \right)^2 = \frac{(\kappa-1)^2+1}{(\kappa-2)^2},
  \label{eqSI:ptpMax_kappa}
\end{align}
with the expected properties $\ptpxMax=1$ for $\kappa=1$ and  $\ptpxMax=1/2$ for $\kappa=0$ when comparing to the analytic results in the inertial and overdamped limits given in the main text.
Also, for $\kappa>0.5$, the summation over unsuccessful events can be disregarded and \cref{eqSI:ptpMax_kappa} can be approximated by $\ptpxMax \approx \kappa^2$.

\subsection{Grote-Hynes result for the transmission coefficient}
\label{SIkappaGH}

According to \citet{Grote1980}, the transmission coefficient $\kappa$ is given by
\begin{align}
    \label{eqSI:kappaGH}
    \kappa = \frac{\lambda}{\OmegaMax},
\end{align}
where $\OmegaMax = \sqrt{|\UPPMax|/m}$ contains the potential curvature at the barrier $\UPPMax$.
$\lambda>0$ is the real reactive frequency, which is the solution of
\begin{equation}
  \label{eqSI:gh_lambda}
  \lambda = \frac{\OmegaMax^2}{\lambda + \tilde{\Gamma}(\lambda) / m },
\end{equation}
where $\tilde{\Gamma}(\lambda)$ denotes the Laplace-transformed memory friction kernel. 

For a single exponential kernel $\Gamma(t) = {\gamma} e^{-t/\tau}/{\tau}$,   $\lambda$ is given as the single positive solution of the cubic equation \cite{Grote1980}
\begin{equation}
\label{eqSI:gh_lambda_exp}
\lambda^3 + \frac{\lambda^2}{\tau} + \left(\frac{\gamma}{m\tau}- \OmegaMax^2 \right) \lambda = \frac{\OmegaMax^2}{\tau}.
\end{equation}
In the Markovian limit, i.e. for delta-correlated friction, $\Gamma(t) = 2 {\gamma} \delta(t)$ and $\tilde \Gamma(\lambda) = {\gamma}$,  
it follows that $\lambda= \left( \gamma^2/(4m^2) +\OmegaMax^2 \right)^{1/2}-\gamma/(2m)$, which is the Kramers medium-to-high-friction result \cite{Kramers1940}.
Note that, either in the inertial, $m\to \infty$, or the long-memory limit, $\tau\to \infty$, it follows that
$\lambda=\OmegaMax$ and thus $\kappa = 1$ in \cref{eqSI:kappaGH}.
The limit  $m\to 0$ is derived by first inserting \cref{eqSI:kappaGH} in  \cref{eqSI:gh_lambda_exp} and expanding $\OmegaMax^2 = |\UPPMax|/m$. By rearranging we find
\begin{align}
\kappa^3 \sqrt{\UPPMax} \tau + \kappa^2 \sqrt{m} + &\kappa \sqrt{\UPPMax} \left(\frac{\gamma}{\UPPMax} - \tau \right) = \sqrt{m},
\end{align}
for which the limit $m\to 0$ leads to
\begin{align}
    \label{eqSI:kappaGH_od}
    \kappa^2 = 1- \frac{\gamma}{\tau \UPPMax}.
\end{align}

For a memory friction kernel of the form $\Gamma(t) = {\gamma_{M}}\delta(t) + \left(\gamma-\gamma_{M}\right) e^{-t/\tau}/{\tau}$, the reactive frequency $\lambda$ is given as the solution of
\begin{equation}
\label{eqSI:gh_lambda_delta_exp}
\lambda^3 + \lambda^2 \left(\frac{1}{\tau} + \frac{\gamma_{M}}{m}\right) + \lambda \left(\frac{\gamma}{m\tau}- \OmegaMax^2 \right) = \frac{\OmegaMax^2}{\tau},
\end{equation}
which leads to the following asymptotic solutions.
For $m \to 0$,
the result is equivalent to \cref{eqSI:kappaGH_od} for single-exponential memory friction.
The results for $m \to \infty$ follows as
\begin{align}
    \kappa^2 = \frac{1}{1+ \tau \gamma_{M}},
\end{align}
and for $\tau \to \infty$ we obtain
\begin{align}
    \kappa^2 = \left(\sqrt{1 + \frac{m \gamma^2_{M}}{4 \UPPMax}} - \sqrt{\frac{m}{\UPPMax}}\frac{\gamma_{M}}{2} \right)^2,
\end{align}
in contrast to the result for single-exponential friction, $\kappa^2 = 1$.
Importantly, these asymptotic results indicate that for $\gamma_{M} \neq 0$, $\ptpxMax \to 1$ is only reached in the double limits $m \to \infty$ and $\tau \to \infty$ or $m \to 0$ and $\tau \to \infty$.

\subsection{Mixed Markovian and non-Markovian friction}
\label{SImixedResults}

\begin{figure}[h]
    \centering
    \begin{overpic}[width=0.4\textwidth]{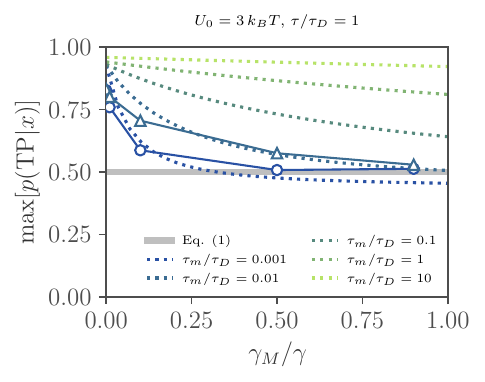}
    \end{overpic}
    \caption{Maximal value of the transition-path probability profiles, $\ptpxMax$, obtained from simulations of the \ac{GLE}, \cref{eq:gle} for $U_0= 3\,k_BT$ and $\tau/\td = 1$, plotted over the ratio of Markovian friction to total friction, $\gamma_{M}/\gamma$, as colored markers for various inertial time scales $\tm/\td$.
    The analytical estimate given by \cref{eq:ptpMax_kappa} in the main text, employing \ac{GH} theory for transmission coefficient $\kappa$, \cref{eq:kappaGH} in the main text, is shown  as colored dotted lines.}
    \label{ptpxMax_gamma}
\end{figure}

The numerical data for mixed Markovian and non-Markovian friction, i.e. $0 < \gamma_{M}/\gamma < 1$, is shown in \cref{ptpxMax_gamma} as colored markers, and compared to the analytical estimate \cref{eq:ptpMax_kappa} in the main text as colored dotted lines.
The analytical estimate successfully captures the decrease of $\ptpxMax$ with increasing $ \gamma_{M}/\gamma$ to the limit value $\ptpxMax=0.5$, which is the well-documented benchmark value for purely Markovian dynamics \cite{Hummer2004}.

